\documentclass[pra,twocolumn,showpacs,floatfix]{revtex4}
\usepackage{graphicx}
\usepackage[usenames]{color}
\newcommand{\beq}{\begin{equation}}
\newcommand{\eeq}{\end{equation}}
\newcommand{\beqa}{\begin{eqnarray}}
\newcommand{\eeqa}{\end{eqnarray}}
\newcommand{\ba}{\begin{array}}
\newcommand{\ea}{\end{array}}

\begin{document}

\title{Spatially antisymmetric localization of matter wave   in a bichromatic
optical lattice}

\author{Yongshan Cheng$^{1,2}$\footnote{yong\_shan@163.com}
and
   S. K. Adhikari$^1$\footnote{adhikari@ift.unesp.br;
URL: www.ift.unesp.br/users/adhikari}}
\affiliation{$^1$Instituto de F\'{\i}sica Te\'orica, UNESP - Universidade
Estadual Paulista,
01.140-070 S\~ao Paulo, S\~ao Paulo, Brazil\\
$^2$Department of Physics, Hubei Normal University, Huangshi 435002, 
People's
   Republic of  China
}

\begin{abstract}

 By direct numerical simulation of the time-dependent
Gross-Pitaevskii equation using the split-step Fourier spectral method
we study the double-humped localization of a cigar-shaped Bose-Einstein
condensate (BEC) in a one-dimensional bichromatic quasi-periodic
optical-lattice potential, as used in a recent experiment on the
localization of a BEC [Roati {\it et al.}, Nature {\bf 453}, 895
(2008)]. {{ Such states are spatially antisymmetric and are excited
modes of Anderson localization.}} Where possible, we have compared the
numerical results with a variational analysis. We also demonstrate the
stability of the localized double-humped BEC states under small
perturbation.

\end{abstract}

\pacs{03.75.Nt,03.75.Lm,64.60.Cn,67.85.Hj }

\maketitle

\section{Introduction}

Fifty years after the prediction of Anderson localization \cite{anderson}
of electron wave
in a disorder potential, the recent experimental
localization \cite{chabe,edwards}
of a non-interacting cigar-shaped
Bose-Einstein condensate (BEC) in a quasi-periodic bichromatic
optical-lattice (OL)
\cite{roati,slava,random,r2,r3,r4,modugno,m2,adhikari,a2,a3,dnlse,d2,d3,d4,d5,d6,d7,d8}
and speckle \cite{billy}
potentials   has
drawn much attention of research workers.
The quasi-periodic bichromatic OL potential \cite{slava}
used in the localization
of a non-interacting BEC \cite{roati}
was formed by the
superposition of two standing-wave polarized laser beams with incommensurate
wavelengths.

Recently, there has been studies of localized BECs with a maximum at the center of the
bichromatic OL trap, where the potential is a minimum \cite{modugno,adhikari,a2,a3}. A natural
extension of this phenomenon would be to investigate localization in more exotic situations,
e.g., where a minimum of the localized BEC is created at the center of the trap. Here, with
numerical simulation of the Gross-Pitaevskii (GP) equation, we study the double-humped
spatially-antisymmetric localization in an excited state \cite{exci,e1,e2,e3,e4,e5} of a
cigar-shaped BEC in a one-dimensional (1D) bichromatic quasi-periodic OL potential. The two humps
appear in a single site of the OL potential. Double-humped structures have been created in a BEC
by phase imprinting and other methods \cite{dark,dr3} and also have been studied theoretically
\cite{thdark,t2,t3,t4}. This makes the present investigation also of experimental interest.
 Anderson localization was originally predicted for the
non-interacting system. However, in the present study on the localization of a
double-humped BEC we also consider a weakly-interacting system.
Although, the present localization is
very similar to Anderson localization in a fully
disordered potential, the  bichromatic OL potential is quasi periodic
and hence deterministic in nature. The localization considered here
is  well described by the 1D discrete Aubry-Andr\'e
model of quasi-periodic confinement \cite{andre,das}.

If the
bichromatic OL potential $V(x)$ has
the symmetry $V(x)=V(-x)$, the localized states $\phi(x)$
of the non-interacting BEC has the symmetry  $\phi(x)=\pm \phi(-x)$.
Here we
consider spatially antisymmetric  localized states
satisfying $\phi(x)=- \phi(-x)$,
which should be considered to be excited modes \cite{exci}
of Anderson localization.

In the presence of strong disorder, the localized state could be quite
similar to a localized state of Gaussian shape in an infinite potential.
However, the more interesting case of localization is in the presence of
a weak disorder when the system is localized due to the quasi-periodic
nature of the potential \cite{roati,billy} and not due to the strength
of the lattice. When this happens the chemical potential of the system
becomes comparable to the height of the bichromatic lattice and the localized
state develops an exponential tail. Nevertheless, the central part of the
localized BEC is found to have a modulated Gaussian shape which allows the
consideration of a variational approximation.

\begin{figure}
\begin{center}
\includegraphics[width=.49\linewidth]{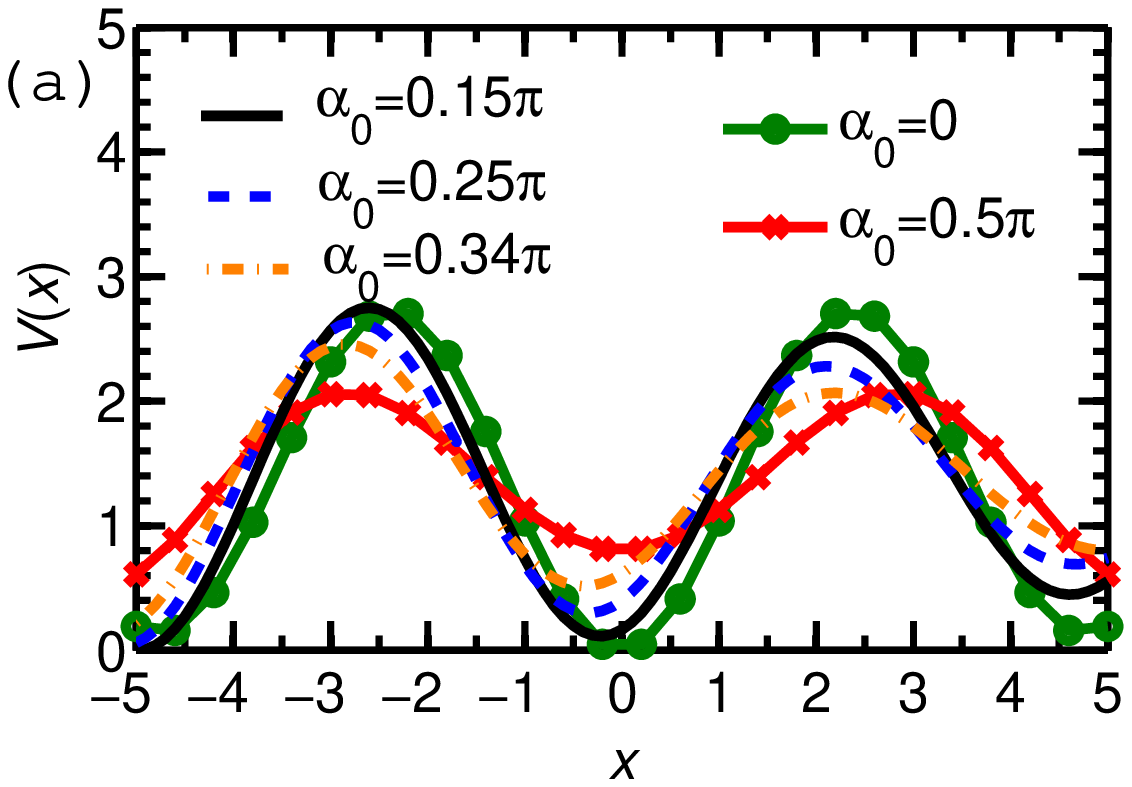}
\includegraphics[width=.49\linewidth]{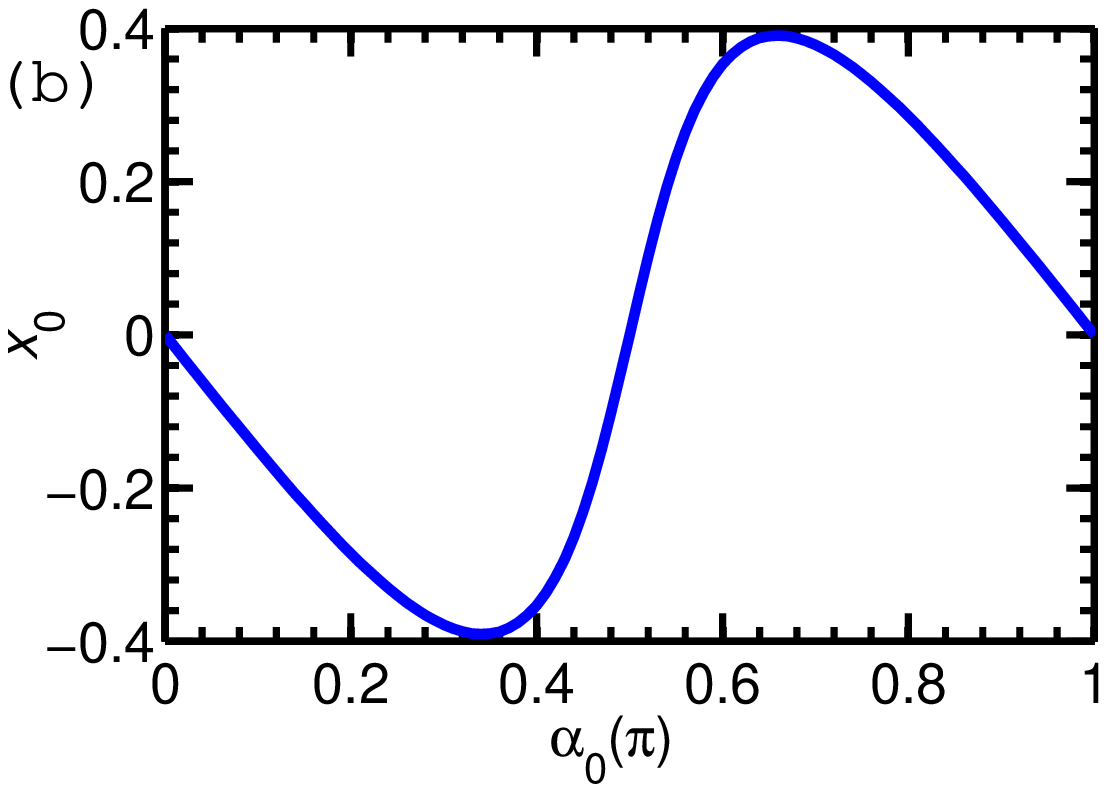}
\end{center}

\caption{(Color online)
        (a) The bichromatic OL potential $V(x)$ vs.  $x$ for different
phases  $\alpha_0$.
 (b) The position $x_0$ of the OL's central minimum
vs.  $\alpha_0$.
Potential parameters in Figs. \ref{fig1} to \ref{fig6} are
$\lambda_1=10, \lambda_2=0.862, \lambda_1, s_1=10$ and $s_2=3$.
}
\label{fig1}
\end{figure}

\section{Analytical consideration}

\label{II}

We consider a cigar-shaped BEC under tight transverse confinement with the bichromatic OL acting
along the axial $x$ direction. Then it is appropriate to consider a 1D reduction
 of the three-dimensional
 GP equation by freezing the transverse dynamics to the respective ground state and integrating
over the transverse variables. The double-humped localized state of $N$ atoms can be described by
the dimensionless GP equation
  \cite{1d,1d2} \begin{equation}\label{eq1} i\frac{\partial u}{\partial t}=-
\frac{1}{2}\frac{\partial^2 u}{\partial x^2} +g|u|^2 u+V(x)u, \end{equation} with normalization
$\int_{-\infty}^{\infty}|u|^2 dx=1,$ of the BEC wave function $u\equiv u(x,t)$.  The spatial
variable $x$, time $t$, and energy are expressed in transverse harmonic oscillator units
$a_\perp=\sqrt{\hbar/(m\omega)}$, $\omega^{-1}$ and $\hbar\omega$, where $m$ is the mass of an
atom and $\omega$ is the angular frequency of the transverse trap with non-linearity \cite{1d}
$g= 2aN/a_\perp^2,$ and $a$ the atomic scattering length. The bichromatic OL potential $V(x)$ is
taken as: \begin{eqnarray}\label{pot} V(x)=\sum_{l=1}^2 A_l \sin^2(k_lx+\alpha_l), \end{eqnarray}
with $A_l=2\pi^2 s_l/\lambda_l^2, (l=1,2)$, where $\lambda_l$'s are the wavelengths of the OL
potentials, $s_l$ are their intensities, $\alpha_l$ are phases, and
  $k_l=2\pi/\lambda_l$ the wave numbers.
We take the phase of the first OL $\alpha_1=0$ and
that of the second
$\alpha_2 \equiv \alpha_0$.
{{In this  investigation,
we take the ratios
$\lambda_2/\lambda_1=0.862$ and
 $s_2/s_1=0.3$ which are roughly the same  as in the
experiment  \cite{roati}.}}
We further take $\lambda_1=10$, and
$s_1=10$.
The experiment of
\cite{roati} employed similar strengths
of the optical lattice to study
Anderson localization
in weak disorder.

For phase difference $\alpha_0=0,$ potential (\ref{pot}) has a minimum at $x=0$.
The position of this minimum moves to $x=x_0\ne 0$ for   $\alpha_0 \ne 0$.
We show in
Fig.  \ref{fig1} (a) the bichromatic OL potential (\ref{pot})
as  $\alpha_0$
is varied.
The bichromatic OL potential is
symmetric around $x=0$
 when $\alpha_0= 0$ or $\pi/2$.
 For the  phase $\alpha_0$ in the range $0<\alpha_0<\pi/2$, the
OL potential is asymmetric, as shown in Fig. \ref{fig1} (a). In Fig.
\ref{fig1} (b) we show the position of the central minimum of potential (\ref{pot})
as  $\alpha_0$ is changed.

\begin{figure}
\begin{center}
\includegraphics[width=0.9\linewidth]{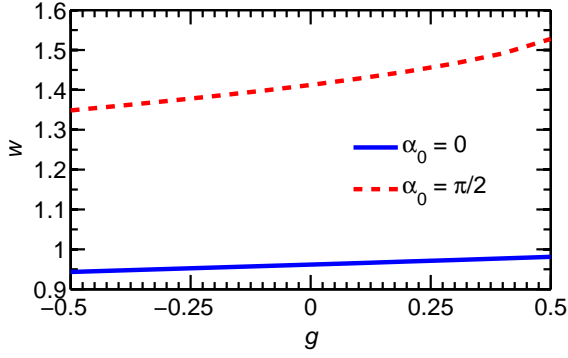}
\end{center}

\caption{(Color online)
The pulse interval $w$ of the localized state vs.  the non-linearity
$g$ for different phase  $\alpha_0$.
}
\label{fig2}
\end{figure}

For a variational analysis of the localized state we consider the stationary wave form $\phi(x)$
given by $u(x, t)=\exp(-i\mu t)\phi(x)$, where $\mu$ is the chemical potential. The real wave
function, $\phi(x)$, obeys the stationary equation, \begin{equation}\label{sta}
\mu\phi(x)+\phi''(x)/2-g\phi^3(x)-V(x)\phi(x)=0, \end{equation} where the prime denotes space
derivative. For $\alpha_0=0$ or $\pi/2$, potential (\ref{pot}) leads to spatially-symmetric or
antisymmetric states confined practically to the central site of the quasi-periodic OL potential.
When this happens, a variational approximation with the following Gauss-type ansatz is useful for
the spatially-antisymmetric state \begin{eqnarray}\label{va}
\phi(x)=\frac{1}{\pi^{1/4}}\sqrt{\frac{{2\cal N}}{w^3}}x \exp\left[-\frac{x^2} {2w^2}\right] ,
\end{eqnarray} where $w$ is the spatial extension of the localized state and will be termed pulse
interval of the localized BEC centered at $x=0$, and ${\cal N}$ is the normalization. The
Lagrangian of the system is given by \begin{eqnarray}
L&=&\int_{-\infty}^{\infty}\left[\mu\phi^2-(\phi ')^2/2-g\phi^4/2 -V(x)\phi^2 \right]dx -\mu ,
\nonumber \\
 &=& \mu {\cal N}-\frac{3{\cal N}}{4w^2}- \frac{3g{\cal
N}^2}{8\sqrt{2\pi}w}-\mu-{\cal N}\sum_{l=1}^2A_l L_l , \\ L_l&=& \left[
\frac{1}{2}+(k_l^2w^2-\frac{1}{2}) \cos(2\alpha_l)\exp(-k_l^2
w^2)\right]. \nonumber \\ \end{eqnarray} The first variational equation
$\partial L/\partial \mu =0$ fixes the normalization: ${\cal N}=1$.
We  use it in the following equations.
The remaining equations $\partial L/\partial w = \partial L/\partial
{\cal N}=0$ yield, respectively, \begin{eqnarray}\label{width}
1&=&\frac{4w^4}{3}\sum_{l=1}^2A_lk_l^2\left( \frac{3}{2} -k_l^2w^2
\right) \cos(2\alpha_l)\exp(-k_l^2w^2)\nonumber \\
&-&\frac{gw}{4\sqrt{2\pi}}, \\
\mu&=&\frac{3}{4w^2}+\frac{3g}{4w\sqrt{2\pi}}+
\sum_{l=1}^2A_lL_l.\label{chem} \end{eqnarray} Equation (\ref{width})
determines the pulse interval of the localized state.
The corresponding energies are given by
$E= \int[(\phi')^2/2+g\phi^4/2+V\phi^2]dx=\mu -3g/(8w\sqrt{2\pi}).$

\begin{figure}
\begin{center}
\includegraphics[width=.49\linewidth]{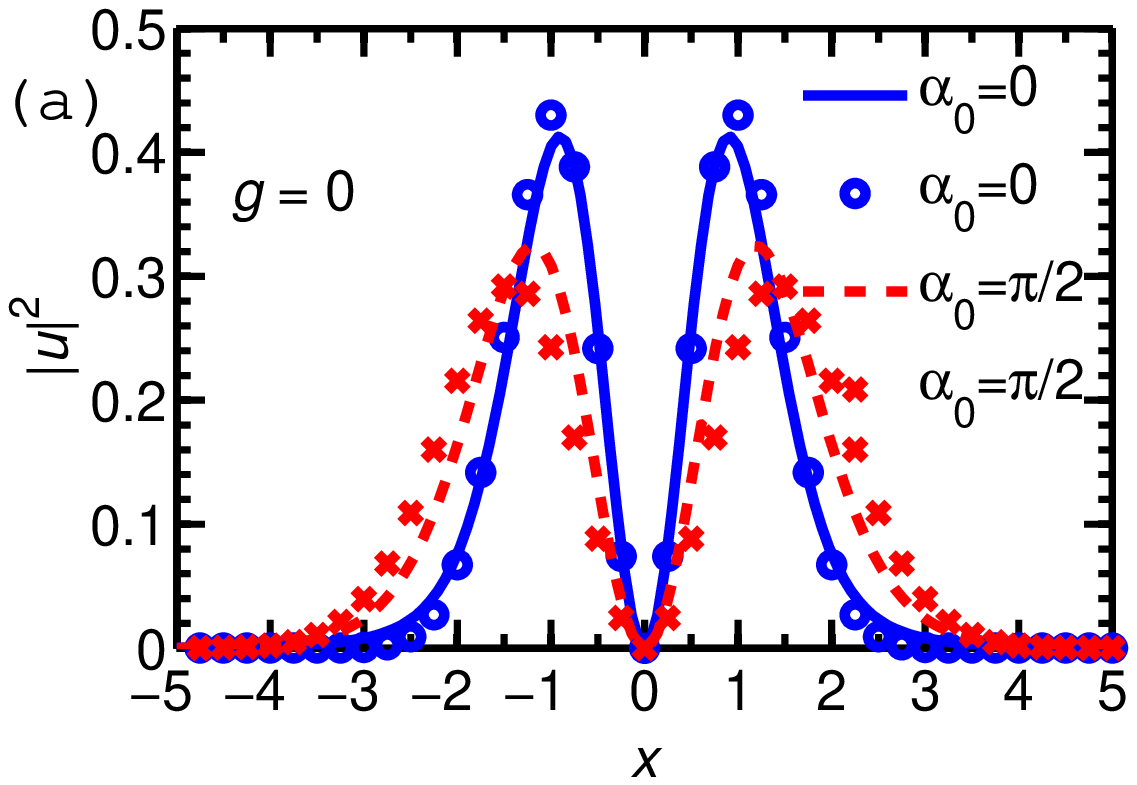}
\includegraphics[width=.49\linewidth]{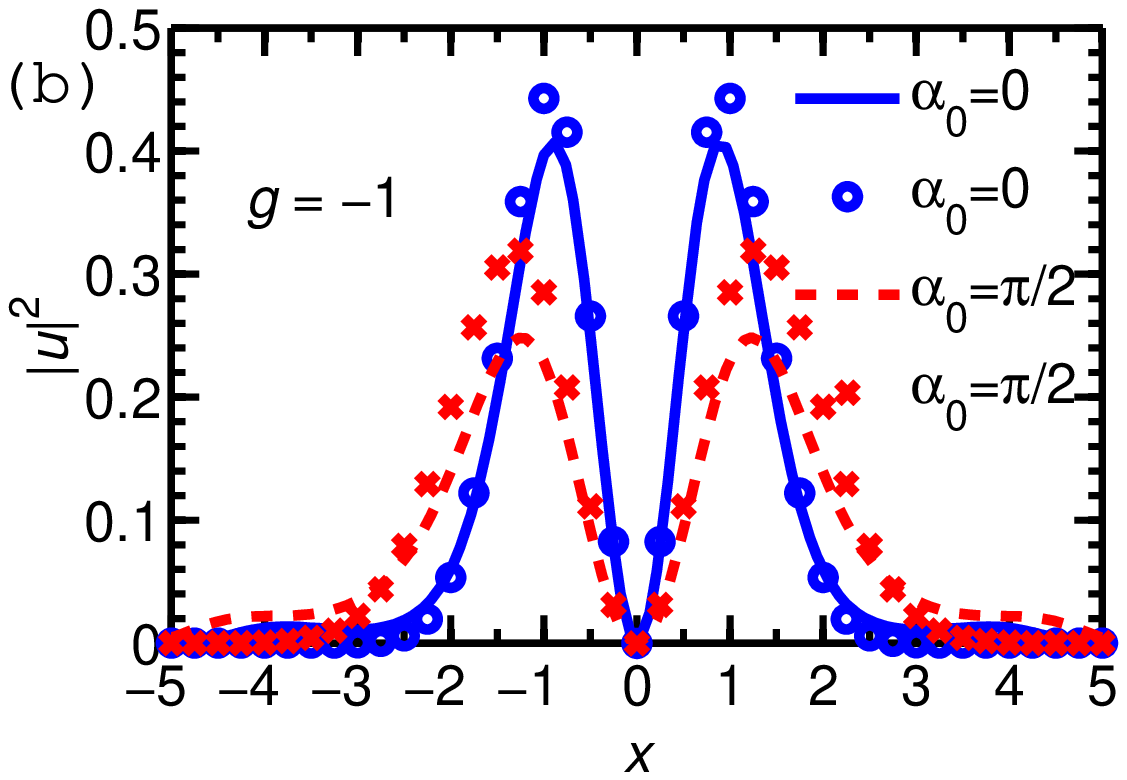}
\includegraphics[width=.49\linewidth]{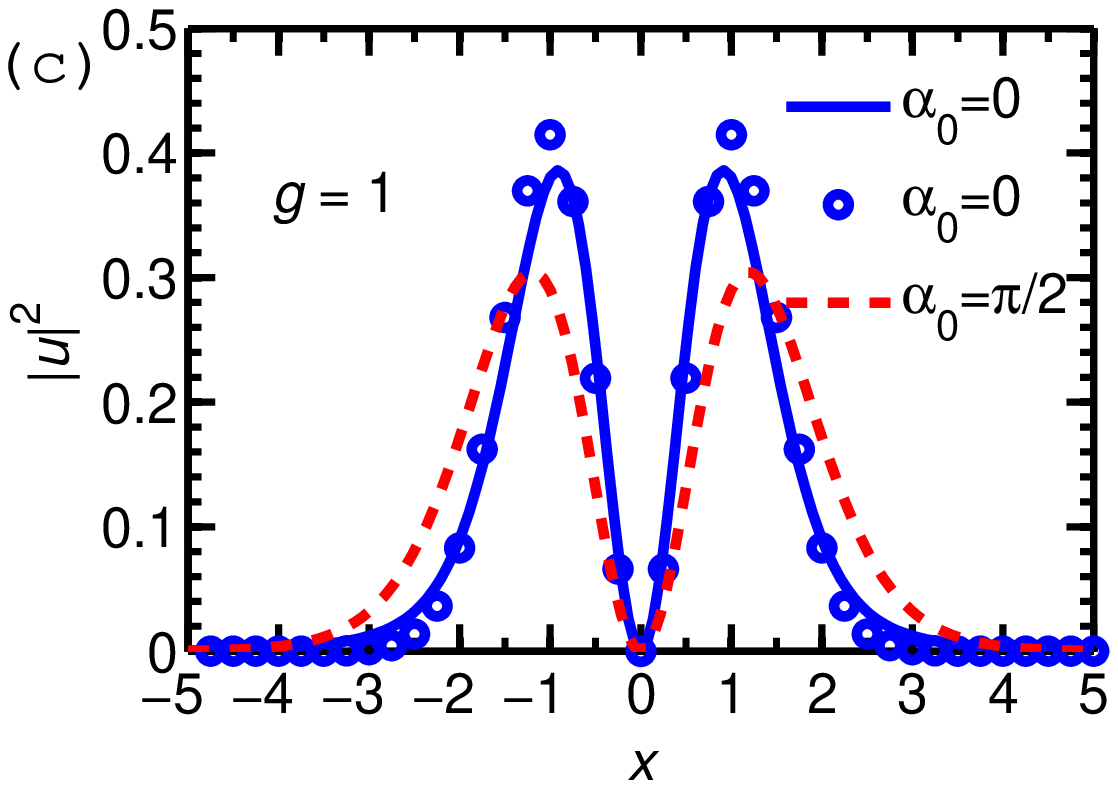}
\includegraphics[width=.49\linewidth]{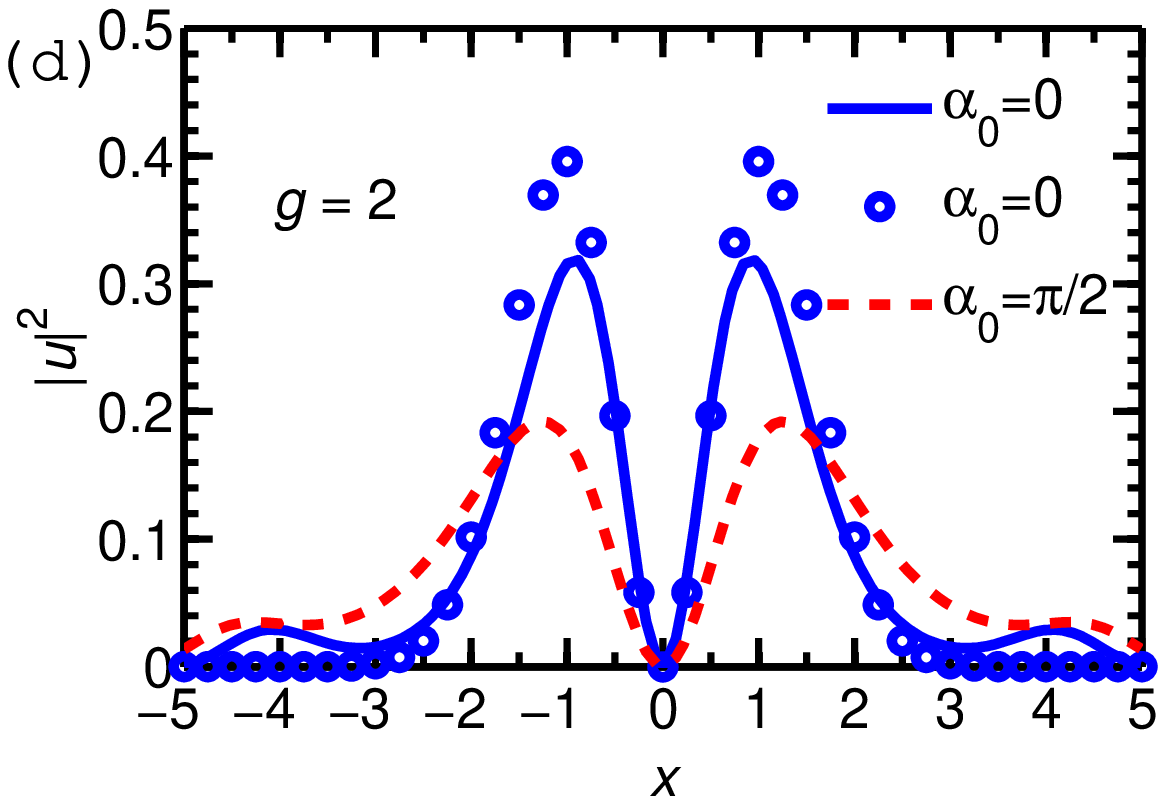}
\end{center}

\caption{(Color online) Numerical (line)
and variational (symbol)
density $|u|^2$ of the antisymmetric double-humped
BEC vs. $x$
for (a) $g=0,$ (b)  $g=-1$,  (c) $g=1$, and (d) $g=2$.
}

\label{fig3}
\end{figure}

\begin{figure}
\begin{center}
\includegraphics[width=.9\linewidth]{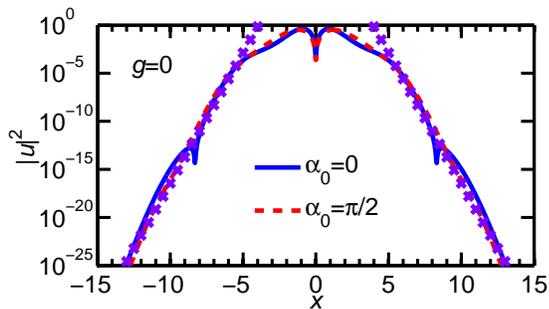}
\end{center}

\caption{(Color online) Numerical  density $|u|^2$ of localized states of
Fig. \ref{fig3} (a) vs. $x$ in log scale. The crosses are exponential
fit $\sim \exp(-|x|/d)$ to density, with localization length $d=0.17$.
}

\label{figx}
\end{figure}

In Fig.
\ref{fig2}, we plot the variational results for the
pulse interval  $w$
vs. the nonlinear coefficient $g$ according to Eq.  (\ref{width}).
The pulse interval
becomes larger as $g$ changes from negative (attractive)
to positive (repulsive). The reason is that
the repulsive interaction among atoms induces a macroscopic repulsion
between the two constituent pulses thus increasing the
pulse interval. Compared with $\alpha_0=0$,
in Fig.  \ref{fig2}, for $\alpha_0=\pi/2$
the pulse
interval is larger and  increases  faster with $g$. This is
because, for $\alpha_0=\pi/2$,
the trapping is weaker relative to $\alpha_0=0$.

\section{Numerical Results}

\label{III}

\begin{figure}
\begin{center}
\includegraphics[width=.9\linewidth]{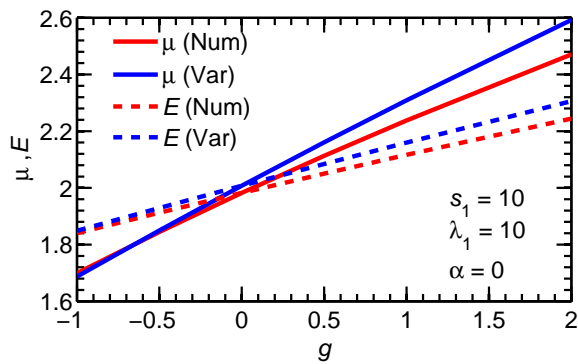}
\end{center}

\caption{(Color online) {{Numerical (Num) and variational (Var) energies
and chemical potential vs. nonlinearity $g$.}}} \label{fig4}
\end{figure}

We perform {the} numerical simulation employing the
real-time split-step
Fourier spectral method  with space step 0.04,
time step 0.0001. Although we use a time-dependent approach the localized
states we calculate are stationary in nature.
The initial input pulse is taken as $u ( x, 0) =\sqrt {2}x\exp(-x^2/2)
/\pi^{1/4}$ with a parabolic trap $V'(x)=x^2/2$ and $g=0$.
In the beginning, during time evolution
the parabolic trap is slowly
turned off and the bichromatic
OL is slowly turned on by increasing $s_1$ by
0.00001 in each time step.
Successively, during time evolution we change   gradually the
nonlinear coefficient $g$ by
0.00001 in each time step to obtain  the stable localized states.

Figures \ref{fig3} (a), (b), (c), and (d)  illustrate
typical  numerical and variational
profiles of the localized states for $g=0,-1, 1$ and 2 for $\alpha_0=0$ and
$\pi/2$.
From Figs. \ref{fig3}
we find  that the numerical
densities  are in good agreement with the variational results
for small non-linearity $g$.
If $g$ is
larger or the trapping is weaker, however,
the localized states develop undulating tails and occupy
more than one OL site.
In that case, density envelope of the localized state can not be
described well by Eq.   (\ref{va}) and the variational approximation will
 no longer be
as good.
We also calculated the chemical potential and energy
of these states using Eqs.  (\ref{width}) and  (\ref{chem}).
The results are shown in Fig. \ref{fig4}.
From Fig.  \ref{fig4} we
find that the energy values are comparable to the trapping potential shown in
Fig. \ref{fig1} (a). This guarantees the interesting limit of weak disorder
as emphasized in  \cite{roati,billy}.

To see this weak disorder explicitly, we plot in Fig. \ref{figx} the
density $|u|^2$, in log scale, of the localized states shown in Fig.
\ref{fig3} (a). The long exponential tail extends from $x=13$ to
$-13$, whereas the central part of density distribution in Fig.
\ref{fig3} (a) contributing to normalization is limited between $x\approx \pm
5$. We have also shown in Fig.  \ref{figx} the exponential fit to
density $\sim \exp(-|x|/d)$ with the localization length
\cite{billy} $d=0.17$.

\begin{figure}
\begin{center}
\includegraphics[width=.49\linewidth]{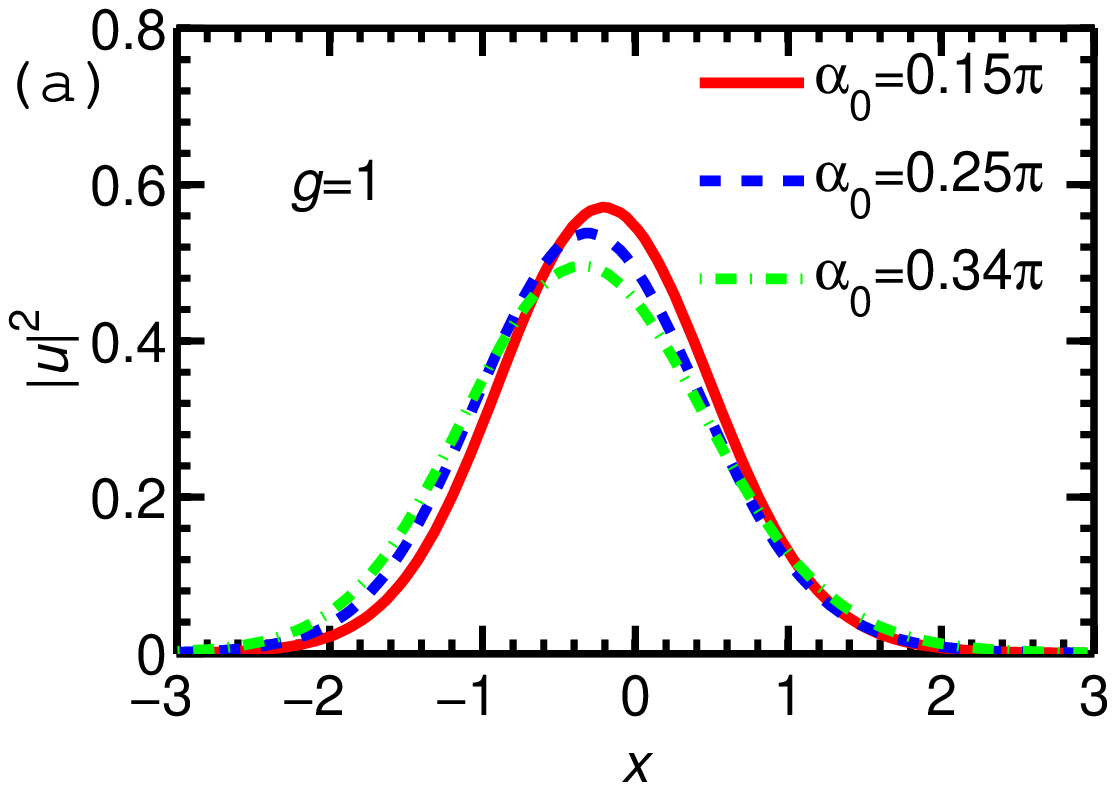}
\includegraphics[width=.49\linewidth]{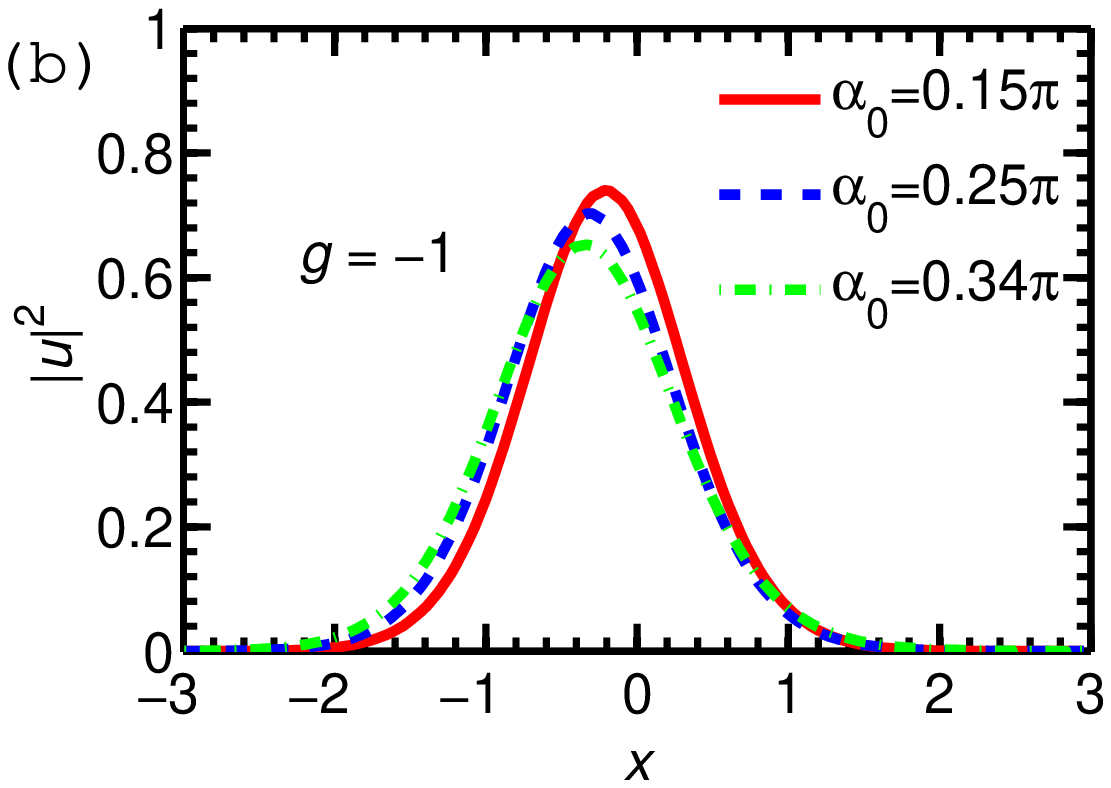}
\includegraphics[width=.49\linewidth]{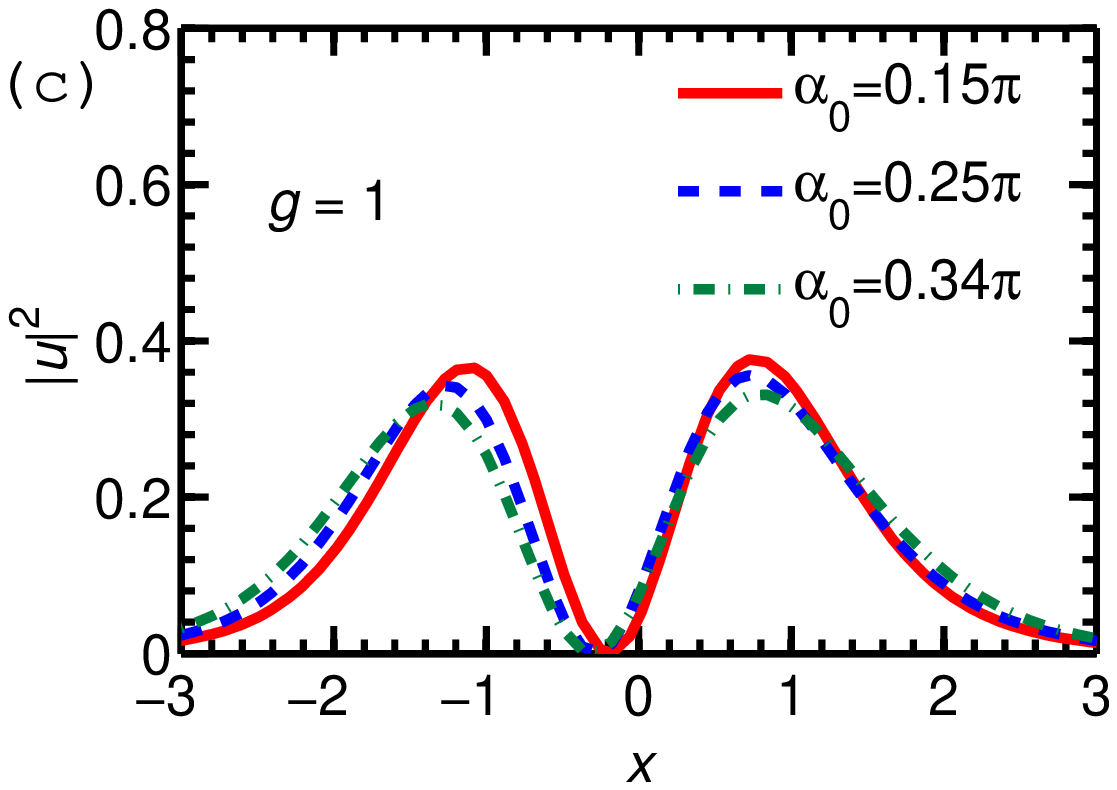}
\includegraphics[width=.49\linewidth]{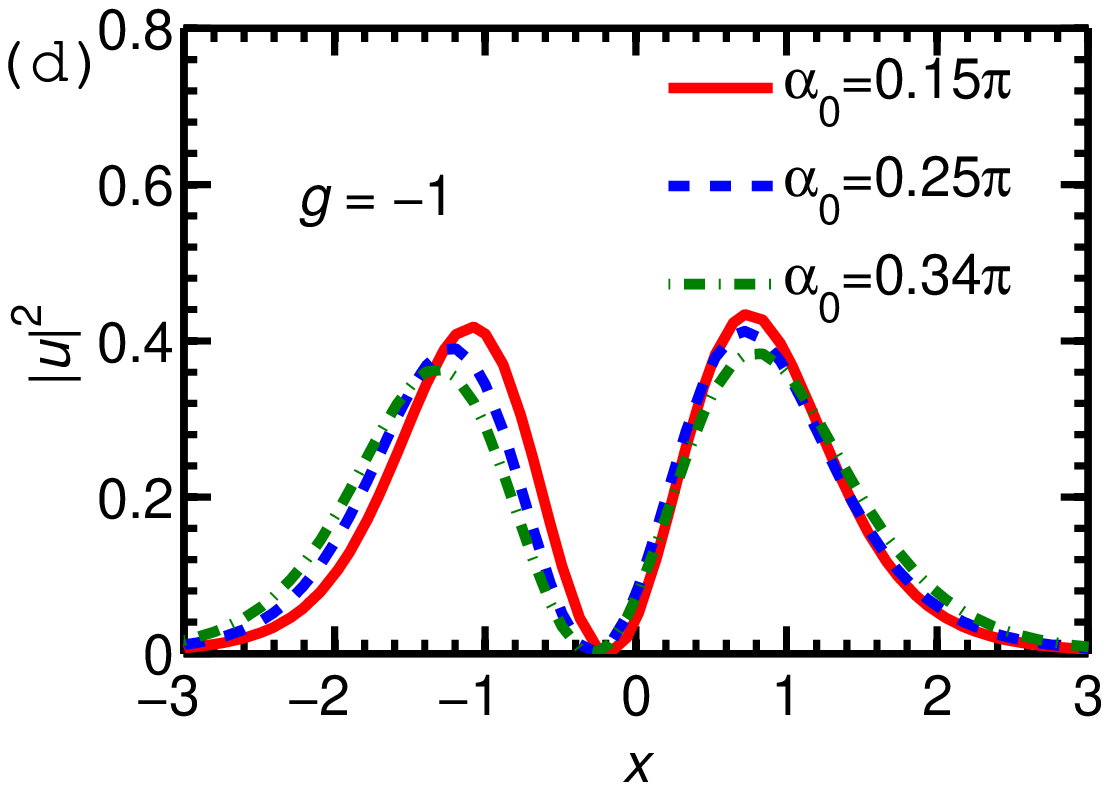}
\end{center}

\caption{(Color online) Numerical density of the asymmetric single-humped
BEC vs.
x for different $\alpha_0$ for (a) $g= 1$ and (b) $g=-1$. (c) and (d)
represent double-humped results for $g=1$ and $-1$, respectively.} \label{fig5}
\end{figure}

Next we present numerical analysis for the asymmetrical OL potential (\ref{pot}) with $\alpha_0
\ne 0$ or $\ne \pi/2$. As shown in Fig. \ref{fig1} (b), the position of the OL's minimum is no
more at $x=0$. On the other hand, the displacement of the OL's minimum changes periodically as
$\alpha_0$ increases. Because of the periodicity, we restrict our attention to
$0<\alpha_0<\pi/2$. In this case, to avoid the loss of accuracy in the localized state in the
numerical integration, {{we move the center of the initial input pulse and the parabolic trap to
the position of the OL's minimum. Thus we calculate the profiles of the localized states in the
asymmetrical OL potential for $g=1,-1$, and $\alpha_0 =0.15\pi, 0.25\pi, $ and $0.34\pi $ and
show the results in Fig.  \ref{fig5}. In order to understand the novel effect, we first
investigate the single-humped localized BECs with the asymmetrical OL potential calculated using
the input pulse $u ( x, 0) = \exp(-x^2/2)/\pi^{1/4}$. The results are shown in Figs.  \ref{fig5}
(a) and (b) for $g=1$ and $ -1$, respectively.
  It can be seen that
the atom density envelopes of the single-hump localized BEC are
asymmetrical.
Compared with Fig. \ref{fig5} (a), Fig.   \ref{fig5} (b)
presents narrower and larger
 atom densities  because of
the focusing nature of the  BEC interaction ($g=-1$).
When $\alpha_0 =0.34\pi
$, the pulse width of the localized state is larger compared to the widths
for
$\alpha_0 =0.15\pi$, or $0.25\pi$. The reason is that the overall
trapping is weaker as $\alpha_0$ increases.

The same results are next  obtained for the double-humped
localized BECs which are shown in Figs. \ref{fig5} (c) and (d)  for $g=1$ and
$g=-1$, respectively, using  the initial input pulse
$u ( x, 0) =\sqrt {2}x\exp(-x^2/2)/\pi^{1/4}$. These two Figs.
show the asymmetry of the two humps for the double-humped localized
BECs. The right hump of the localized states is higher than its left
hump because trapping on the left is stronger. When $\alpha_0
=0.34\pi $, the difference in the height of the two humps and the
pulse interval of the localized state are larger compared with those of
$\alpha_0=0.25\pi$ or $0.15\pi$. These results are in agreement with
single-humped localized BECs illustrated in Figs. \ref{fig5} (a) and (b). }}

\begin{figure}
\begin{center}
\includegraphics[width=.49\linewidth]{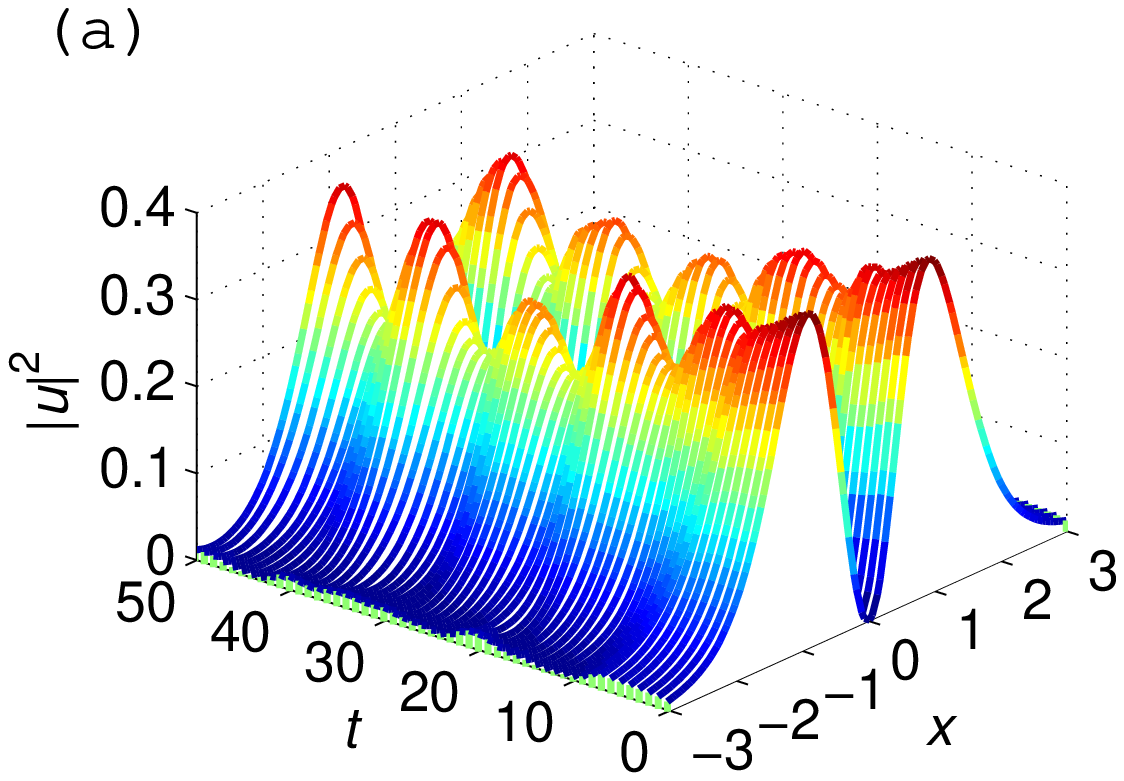}
\includegraphics[width=.49\linewidth]{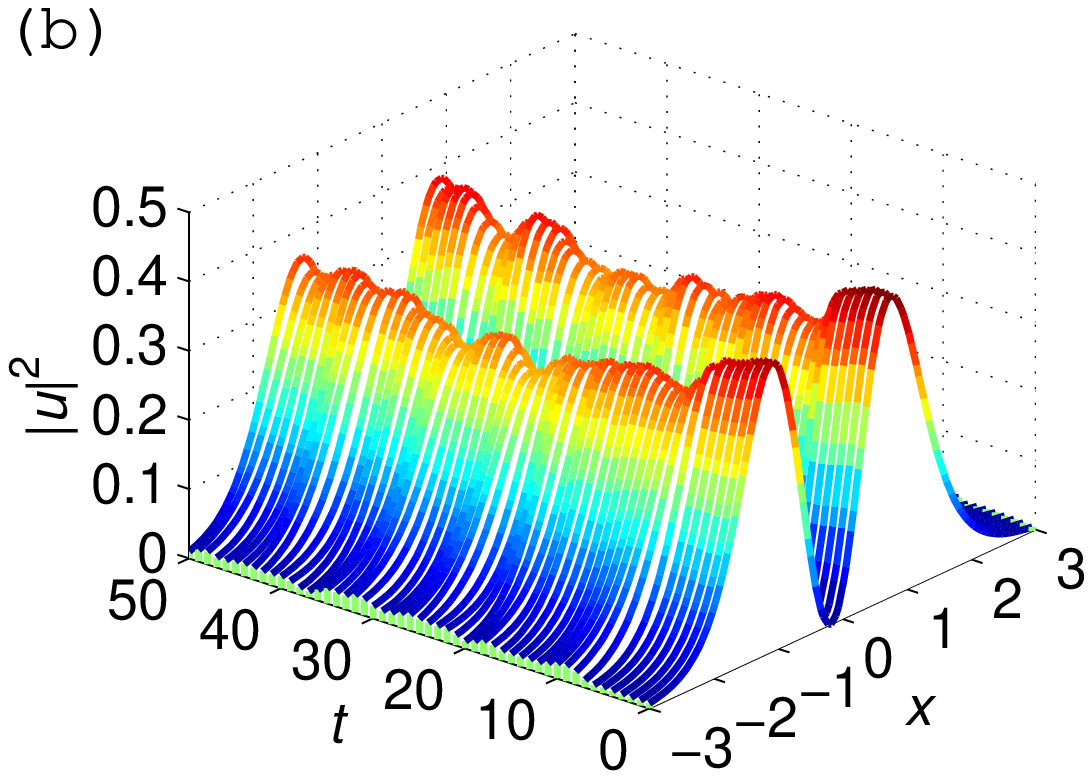}
\end{center}

\caption{(Color online) Stability dynamics $|u(x,t)|^2$ vs. $x$ and $t$
for (a) the antisymmetric state with $g=1, \alpha_0=0, $ [viz.  Fig. \ref{fig3}
(c)] as the center of the OL trap is displaced by a small distance
$\Delta x=0.1$ at time $t=5$, and for (b) the asymmetric state with $g=-1,
\alpha_0=0.15\pi $ [viz. Fig. \ref{fig5} (d)]  as the wave length $\lambda_1$
is suddenly changed from 10 to 11 at time $t=5$.
}
\label{fig6}
\end{figure}

\begin{figure}
\begin{center}
\includegraphics[width=0.9\linewidth]{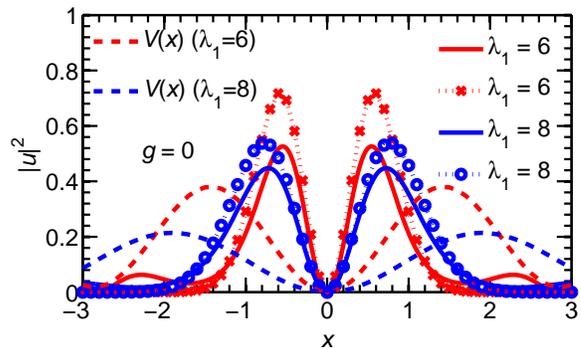}
\end{center}

\caption{(Color online) The numerical (solid line) and variational
(dotted line with symbol)
density   $|u|^2$
vs. $x$ of the double-humped BEC  for $\lambda_1=6$ and 8, and
$g=0, s_1=10, s_2=3,
\lambda_2=0.862 \lambda_1$, and $\alpha_0=0$.
The corresponding bichromatic OL potential is also shown
in arbitrary units.
}
\label{fig7}
\end{figure}

One of the most important issues is the stability of the double-humped localization against
perturbation. First, we investigate the stability of the antisymmetrical localization. In order
to examine whether the predicted localized state is stable, we introduce a small perturbation
from equilibrium point by displacing the center of the OL trap by a small distance $\Delta x=0.1$
at $ t=5.$
 The numerical simulations are shown in Fig.  \ref{fig6} (a)
 for $g=1, \alpha_0=0$. The double-humped localization  oscillate
around the new center of the OL trap, and the symmetry of the two humps
is broken after the small perturbation is introduced.
The double-humped
localization exists during a large  time interval for small
perturbation. We also checked the stability of the states for other values
(negative)
of $g$  and for other types of perturbation by changing the value of the
wave length $\lambda_1$
 and the antisymmetrical localized state was found to be stable.
Next we studied the stability of the asymmetric localized states.
The stability of one such state is shown in Fig.  \ref{fig6} (b) for
$\alpha_0=0.15\pi$ and $g=-1$
when $\lambda_1$ was suddenly changed from 10 to 11
at time $t=5$.
The double-humped state is again found to be stable against small perturbation.

So far we considered a fixed  OL wave length: $\lambda_1=10$.
To get
insight into the effects of the OL on the stationary localized states,
we now investigate the properties of the system when the wave lengths
$\lambda_l$ are smaller. In
realistic experiment, the dimensionless wavelength can be adjusted by
the harmonic trap because $\lambda_1$  is related to the transverse
harmonic-oscillator length. The wave length
$\lambda_1$ controls
the dimensionless
heights of potential (\ref{pot}),
so it is meaningful to
investigate the effects of $\lambda_1$
on the stationary localized states. In
this part, the same parameters of the numerical integration are selected
as those in the preceding calculation, (viz.
$ s_1=10, s_2=3$,  $\lambda_2=0.862\lambda_1$, and $\alpha_0=0$).
In Fig.  \ref{fig7}
we plot $|u|^2$ vs. $x$
for $g=0$ and     different $\lambda_1$.
A smaller $\lambda_1$
 leads to a  localized state with a smaller pulse interval $w$.
{
We have also compared the numerical results with variational analysis
for small $\lambda_1$. The variational results are shown by
the chain of symbols in Fig.  \ref{fig7}.
}

\section{SUMMARY}

Using the numerical and variational solution of the
 GP equation, we studied the stationary localization of a
double-humped cigar-shaped BEC in a bichromatic
quasi-periodic 1D OL potential.
The bichromatic OL potential is generated by
superposing two OL potentials in the form
of sine waves.  Such a bichromatic OL potential is symmetric around the
center at $x=0$, consequently, the density of the localized state also
possesses the same symmetry. In the presence of a phase difference
between the two OL components the above symmetry is broken and we
analyze this symmetry breaking in case of single-humped and
double-humped states.
Here we also  study the effect of a weak atomic interaction (both attractive
and repulsive)
on the profile of the
localized states.
The localized double-humped states were found to be dynamically stable
under small perturbations.
We hope that the present work will motivate new studies, specially
experimental ones on Anderson localization in the form of
double-humped states.

\label{IIII}

\acknowledgments

FAPESP and CNPq (Brazil) provided partial support.

\end{document}